# A Closed Algebra of Clebsch Forms Derived from Whittaker Super-potentials and applications in electromagnetic research

T. E. Raptis

Division of Applied Technologies

National Center for Science and Research, "Demokritos"

Email: *rtheo@dat.demokritos.gr*

**Abstract:** A type of closed exterior algebra in $R^3$ under the cross product is revealed to hold between differential forms from the three Whittaker scalar potentials, associated with the fields of a moving electron. A special algebraic structure is revealed in the context of Clebsch reparametrization of these scalars, and a special prescription for the construction of permutation invariant electromagnetic fields is given as well as a superposition with parallel electric and magnetic components.

1. Introduction

At 1904, Edmund Whittaker [1], based on a well known formula for the integral solutions of wave equations by his former student Harry Bateman [2], discovered that the overall electromagnetic field produced by a moving electron in retarded coordinates, can be analyzed in three scalar potential terms. These were showed to be able to reproduce the vector potential of the moving electron, thus been cast later into the generic class of so called, Superpotentials.

Later on, H. S. Ruse showed in two subsequent papers [3] [4] the intrinsic geometric significance of these functions as the principal directions of a covariant relativistic expression of Hertzian tensors in general. Furthermore, Kawaguchi in a series of papers [5 - 9], emphasized the significance of such superpotentials starting from a Hertzian analysis of the Lienard-Wiechert potentials in the effort of discovering an underlying geometrization of Maxwellian electrodynamics.

In what follows we intend to provide examples of another source of such superpotentials from a different decomposition of vector potentials that has been overlooked by modern engineering practice in electromagnetism. This was fairly well known in hydrodynamicists and one can take this link further by noticing the important correspondence of Maxwell's theory with the theory of the ideal Euler fluid [10], [11]. The initial Whittaker scalars will be used as a source of a particular algebraic structure which then gives rise to unusual configurations of the electromagnetic field. These might prove important in certain areas of application like pulsed power, "streamers", discharges and lightning phenomena.

## 2. The Clebsch representations of arbitrary solenoidal vector potentials

The generic representation of arbitrary vector fields by Clebsch scalar potentials, originated in 19$^{th}$ century hydrodynamics, in the effort to classify the various solutions of the generally non-linear Navier-Stokes equations and their many possible reductions into special cases [12 - 16]. It has since been used extensively only in various specialized areas, like Magneto-hydrodynamics, Heliodynamics, Geophysics and plasma physics. As such, its possible uses in Maxwellian electrodynamics have been largely suppressed and the technique is not very popular in a more general curriculum in the physics community. Most of all, it's non-uniqueness makes difficult the applicability of the method while it's connection with the general Helmholtz theory is not straightforward and in fact is a non-linear one as we discuss in this section.

The Clebsch parametrization consists of the following very general representation of arbitrary vector fields with three scalars in the form

(1) $\mathbf{F} = \nabla\varphi + \nabla\times(\xi\nabla\psi) = \nabla\varphi + \nabla\xi\times\nabla\psi$

In hydrodynamics literature, the two last scalars $\{\xi,\psi\}$ are often referred to as the "stream" and "flux" potentials. The above representation is in a direct correspondence with the generic Helmholtz decomposition usually given as

(2) $\mathbf{F} = \nabla\varphi + \nabla\times\mathbf{A}$

Apparently, the correspondence is not unique because of the inherent non-linearity of the Clebsch representation. This can be shown by writing the direct relationship of the initial field *F* and the stream and flux potentials as

(3) $\nabla\times\mathbf{F} = \nabla\times(\nabla\xi\times\nabla\psi)$

Otherwise, if the vector potential can be found by a direct use of the inverse curl or "*Biot-Savart*" integral operator, the correspondence is given directly as $\mathbf{A} = \nabla\xi\times\nabla\psi$. Such correspondence though, results into a system of three coupled nonlinear PDEs of which no general solution can be given in a unique form. This makes Clebsch forms more difficult for everyday engineering practice and is perhaps one of the reasons it has been overlooked.

Yet, there is one case where the introduction of such scalars may adopt a special meaning. It is the case of super-potentials as irreducible representations of an arbitrary field theory and its deeper geometrical structure. Super-potentials have for long been discussed in association with a possible complete geometrization of electrodynamics [8]. In fact, the apparent plurality of possible Clebsch representations of 4-vector potentials creates a unique problem on the types of symmetries hidden in some functional group structures. As far as the author knows, such a problem has not been exhaustively searched.

We introduce the technique of Clebsch decomposition by making use of the Whittaker scalars due to an unexpected symmetry contained in them. These were introduced in [1] as a triplet of scalars $\{F, G, \psi\}$ from which the corresponding electric and magnetic fields of an electron moving in the z direction in retarded coordinates could be reconstructed in the symmetric forms

$$
(4) \quad \begin{aligned}
\mathbf{E} &= \frac{1}{c}\nabla \times [(\partial_t G)\mathbf{z}] + \nabla \times \nabla \times (F\mathbf{z}) \\
\mathbf{B} &= \frac{1}{c}\nabla \times [(\partial_t F)\mathbf{z}] - \nabla \times \nabla \times (G\mathbf{z})
\end{aligned}
$$

In the above, c is as usual the speed of light and E and B the electric and magnetic vectors in free space. The particular scalars where also expressed in a simplified form by Ruse [4] as follows

$$
(5) \quad \begin{aligned}
F &= \tanh^{-1}(\frac{z-z_0}{t-t_0}) \\
G &= \tan^{-1}(\frac{y-y_0}{x-x_0}) \\
\psi &= \frac{1}{2}\ln\left((x-x_0)^2 + (y-y_0)^2\right)
\end{aligned}
$$

In the original work of Whittaker and Ruse, these are used to derive expressions for the scalar and vector electromagnetic potentials. When introduced in the defining relations for the electric and magnetic vectors, the $\psi$ scalar disappears in the final expressions given by (4).

In the next section, we make a different use of the Whittaker scalars as possible abstract generators of new solutions to Maxwell equations via the Clebsch technique. Specifically, we treat all $\binom{2}{3}$ combinations of pairs as possible Clebsch potentials for the direct construction of electric and magnetic vectors and we study the particularly simple and elegant algebraic structure that they obey. It should be emphasized that the use of the particular scalars is not given here as an alternative for the original fields of a moving electron but rather as a special case of a general technique for deriving other types of field configurations. It is only with respect to the particular geometric structure that occurs from differentiation of the above that this choice is justified here. The case of more general functions is discussed in the last section.

3. **The exterior algebra of forms derived from Whittaker scalars**

At this section it is shown that the original Whittaker scalars have some unusual properties if seen as functions of stationary coordinates $(x, y, z, t)$, such that they allow the construction of an entirely new class of electromagnetic fields. We first rewrite them in the form

$$(6) \quad \begin{aligned} F &= \tanh^{-1}(u(z,t)) \\ G &= \tan^{-1}(v(x,y)) \\ \psi &= \frac{1}{2}\ln(|\mathbf{k}_\psi|^2) \end{aligned}$$

In the above we have introduced the auxiliary variables

$$(7) \quad \begin{aligned} u(z,t) &= \frac{z-z_0}{t-t_0} \\ v(x,y) &= \frac{y-y_0}{x-x_0} \\ \mathbf{k}_\psi &= (x-x_0, y-y_0, 0) \end{aligned}$$

We then immediately derive

$$(8) \quad \begin{aligned} \nabla F &= \frac{1}{1-u^2}(\partial_z u)\mathbf{z} = \frac{1}{(1-u^2)(t-t_0)}\mathbf{z} \\ \nabla G &= \frac{1}{1+v^2}\nabla_{(x,y)}v = \frac{1}{(1+v^2)(x-x_0)}(-v \ \ 1 \ \ 0) \\ \nabla \psi &= \frac{1}{2|\mathbf{k}_\psi|^2}\nabla_{(x,y)}|\mathbf{k}_\psi|^2 = \frac{1}{\|\mathbf{k}_\psi\|^2}\mathbf{k}_\psi \end{aligned}$$

We then see that all derivatives can be re-expressed with the aid of three characteristic vectors as

$$(9) \quad \begin{aligned} \nabla F &= A(z,t)\mathbf{k}_F \\ \nabla G &= B(x,y)\mathbf{k}_G \\ \nabla \psi &= \Gamma(x,y)\mathbf{k}_\psi \end{aligned}$$

These are then given as

$$\mathbf{k}_F = \begin{pmatrix} 0 & 0 & 1 \end{pmatrix}^T$$

(10) $$\mathbf{k}_G = \begin{pmatrix} -(y-y_0) & x-x0 & 0 \end{pmatrix}^T$$

$$\mathbf{k}_\psi = \begin{pmatrix} x-x_0 & y-y_0 & 0 \end{pmatrix}$$

The particular pairs $\{\mathbf{k}_F, \mathbf{k}_G\}$ and $\{\mathbf{k}_F, \mathbf{k}_\psi\}$ have a structure resembling what is known in engineering electromagnetism as a TE/TM pair. The two last also have the additional properties $\nabla \bullet \mathbf{k}_G = 0, \nabla \bullet \mathbf{k}_\psi = 2, \nabla \times \mathbf{k}_G = \nabla \times \mathbf{k}_\psi = 0$ while they are connected via a linear transformation given by the 3 x 3 matrix.

(11) $$\mathbf{M}_J = \begin{pmatrix} \mathbf{J} & 0 \\ 0 & 1 \end{pmatrix}, \quad \mathbf{k}_G = \mathbf{M}_J \bullet \mathbf{k}_\psi$$

In the above we have used the 2 x 2 symplectic "*Darboux*" matrix

(12) $$\mathbf{J} = \begin{pmatrix} 0 & -1 \\ 1 & 0 \end{pmatrix}, \quad \mathbf{J}^2 = -\mathbf{I}$$

It is then possible to write the closed algebra of these vectors in the form

(13)
$$\mathbf{k}_F \times \mathbf{k}_G = -\mathbf{k}_\psi$$
$$\mathbf{k}_F \times \mathbf{k}_\psi = \mathbf{k}_G$$
$$\mathbf{k}_G \times \mathbf{k}_\psi = \gamma(x,y)\mathbf{k}_F, \quad \gamma = (x-x_0)(v^2-1)$$

If we now construct the three vector fields of the associated Clebsch forms $\mathbf{B}_1 = \nabla F \times \nabla G, \mathbf{B}_2 = \nabla F \times \nabla \psi, \mathbf{B}_3 = \nabla G \times \nabla \psi$, these can be reduced back to the simpler forms $\mathbf{B}_1 = \lambda_1 \nabla \psi, \mathbf{B}_2 = \lambda_2 \nabla, \mathbf{B}_3 = \lambda_3 \nabla F$ where $\lambda_i$ appropriate proportionality scalar functions.

We see that the above field has the character of a time-dependent magnetic field as every Clebsch form is irrotational. For simplicity we write directly its explicit dependence for easier use in what follows as

$$\mathbf{B} = \alpha \mathbf{k}_F - \beta \mathbf{k}_G + \gamma \mathbf{k}_\psi$$

(14) $$\alpha = \frac{v^2-1}{v^2+1}, \quad \beta = \frac{t-t_0}{|\mathbf{k}_Z|^2|\mathbf{k}_\psi|^2}, \quad \gamma = \frac{(x-x_0)(t-t_0)|}{|\mathbf{k}_Z|^2|\mathbf{k}_\psi|^4}$$

$$\mathbf{k}_Z = (z-z_0, \quad \mathbf{i}(t-t_0))$$

At this point we emphasize that the parameters $(x_0^i, t_0)$ are not of particular importance in the subsequent derivation and can be omitted in the final result but we kept them for generality. The poles appearing near the origin limit their validity outside of the appropriate boundary conditions that should be put on the sources.

Next, we check whether we could associate with the above field a complementary electric field given that the following equations can have at least one solution

$$(15) \quad \begin{aligned} \nabla \times \mathbf{E} &= -\partial_t \mathbf{B} = -(\partial_t \beta)\mathbf{k}_G - (\partial_t \gamma)\mathbf{k}_\psi \\ \frac{1}{c^2}\partial_t \mathbf{E} &= \nabla \times \mathbf{B} - \mu_0 \mathbf{j} \\ \nabla \bullet \mathbf{E} &= \rho/\varepsilon_0 \\ \nabla \bullet \mathbf{j} &= \partial_t \rho \end{aligned}$$

Since the first of the above is an independent relation, we can take twice the curl to end up with a Poisson equation

$$(16) \quad \nabla^2 \mathbf{E} = \frac{1}{\varepsilon_0}\nabla\rho + \nabla(\partial_t \beta)\times \mathbf{k}_G - \nabla(\partial_t \gamma)\times \mathbf{k}_\psi$$

We can also find the appropriate current directly from the second equation in the form

$$(17) \quad \mathbf{j} = \frac{1}{\mu_0}\nabla \times \mathbf{B} - \varepsilon_0 \partial_t \mathbf{E}$$

To ease the construction of the electric part we consider again the closed algebra of the characteristic vectors. In particular for the vorticity of the electric field to have a projection solely into the last two vectors we may choose a Clebsch parametrization of the form $\mathbf{E} = \nabla\Phi + g_1 \nabla h_1 + g_2 \nabla h_2$. Then the first of (15) becomes

$$(18)\; \nabla g_1 \times \nabla h_1 + \nabla g_2 \times \nabla h_2 = -(\partial_t \beta)\mathbf{k}_G - (\partial_t \gamma)\mathbf{k}_\psi$$

The above can be greatly simplified by a reduction of the unkown functions which is possible with the introduction of two arbitrary vectors such that $\nabla g_i = \mathbf{k}_i = \begin{pmatrix} k_x^i() & k_y^i() & 0 \end{pmatrix}$. Then, the two pairs of unknowns can be separated to give the conditions

(19)
$$\nabla h_1 \times \mathbf{k}_1 = (\partial_t \beta)\mathbf{k}_G$$
$$\nabla h_2 \times \mathbf{k}_2 = (\partial_t \gamma)\mathbf{k}_\psi$$

From the resulting linear system we get the final conditions in the form

(19)
$$\partial_z h_i = -\left(\frac{x-x_0}{\partial_x g_i}\right)\partial_t \beta = -\left(\frac{y-y_0}{\partial_y g_i}\right)\partial_t \beta$$
$$\partial_y h_i - v\partial_x h_i = 0$$
$$\partial_y g_i - v\partial_x g_i = 0$$

The above conditions reveal a whole class of possible solutions. Last but not least, we mention that it is possible to create such a solution with a solenoidal current satisfying $\nabla \bullet \mathbf{J} = 0$ for a static charge distribution if a class of functions $g_i, h_i$ can be found that satisfies $\partial_t \nabla \bullet \mathbf{E} = 0$. This is then equivalent to the symmetry condition $\partial_t \nabla \bullet (g_1 \nabla h_1) = -\partial_t \nabla \bullet (g_2 \nabla h_2)$ which should be added to (19) to constraint the four initial degrees of freedom represented by the scalars $\{g_i, h_i\}$.

Next, we expose the most important properties of the fields constructed from the above prescribed Clebsch algebras, which is their invariance under permutations and we discuss their possible physical significance.

4. **Discussion and conclusions**

In the previous section, it was shown that there exist closed distributions of charges and current volumes that give rise to a peculiar class of quasi-static electromagnetic fields. The particular class appears to be a hybrid mode, composed of a static longtitudinal near field in the z-direction, while it has a pair of hyperbolically varying components in both spatial and temporal dimensions, resembling a "leakage" field which could be akin to plasma discharge and lightning phenomena. Additionally, one can find areas of applications in near field optics where such components appear to be of importance as in the case of inductive energy transfer as well as in microcircuits inductive coupling.

It appears that due to the freedom in the choice of the Clebsch superpotentials for the electric field, one can find situations where a closed current distribution could be used to form and sustain such electromagnetic configurations in the form of slowly decaying pulses or "streamers", a fact that makes the above attractive for the case of the so called, "*ball lightning*". One should also notice the possibility of such current distributions to be attributed solely on a particular type of internal polarization and/or magnetization, assuming a linear, non-isotropic material. It is of course entirely possible to generalize the previously found configurations with a little more effort into the context of Clebsch analysis for generally anisotropic and inhomogeneous materials but this goes beyond the scope of the present short report.

In order to completely cover the issue of the particular symmetry involved in the choice of the initial superpotentials from the Whittaker scalars, one should proceed to a closer examination of its symmetry properties. In particular, the characteristic separation between the $(x, y)$ and $(z, t)$ spatial degrees of freedom is akin to permutations that leave the vector potential in (4) invariant with only modifying the unit vector to a new direction. Hence, if we use a symbolic operator $\hat{\pi} : \hat{\pi}\{(x,y),(z,t)\} \to \{(y,z),(x,t)\}$ then the action of this on the vector potential will just be a shift of the unit vector $\hat{\mathbf{z}} \to \hat{\mathbf{x}}$ in (4).

This property is generic and holds true even for the electromagnetic fields produced from the Clebsch decomposition of section **3.** That is, if one denotes the pair of resulting fields with the symbolic notation $\{\mathbf{E},\mathbf{B}\}(x,y;z,t)$ then the action of the $\pi$ operator on their arguments will produce equally valid fields. This action would then produce permutations of the elements of the three characteristic vectors $\{\mathbf{k}_F, \mathbf{k}_G, \mathbf{k}_\psi\}$ involving exchange of pairs as implied by the first part of the action of the $\hat{\pi}$ operator. Hence, one ends with a triplet of possible electromagnetic configurations differently oriented.

Apparently, this allows the construction of a special superposition of all three of them in the form of total fields

(20) $\quad \{\mathbf{E}',\mathbf{B}'\} = \sum_{i=1}^{3} \{\mathbf{E}^{(i)}, \mathbf{B}^{(j)}\}\left(\hat{\pi}^i(x_i, x_{i+1}; x_{i+2}, t)\right)$

If we check the electromagnetic invariant $\mathbf{E} \bullet \mathbf{B}$ for such fields we see that taking into account all terms it depends on 27 products given as $\left\{\mathbf{k}_i^j \bullet \nabla \Phi, \mathbf{k}_i^j \bullet \nabla h_{1,2}\right\}_{\substack{i \in \{F,G,\psi\} \\ j=1..3}}$. In general, we are dealing with non-transverse or as is otherwise known, non-null radiation. To see the exact nature of this geometry for the total superposition, we express each term in the Cartesian system. With the aid of (10) and (14) this takes the simple form

(21) $\quad \mathbf{B}^{(i)} = \left(\hat{\pi}^i(\gamma x + \beta y), \quad \hat{\pi}^i(\gamma y - \beta x), \quad \hat{\pi}^i(\alpha)\right)$

Whatever the angle with the corresponding electric vector *E*, the three angles corresponding to the three exchanges of coordinates, will necessarily produce a symmetric arrangement of them around the diagonal of the Cartesian frame with the three total vectors for each total field making a symmetric tetrahedron with its major apex at the origin of the Cartesian frame and its base formed from the edges of the three total electric or magnetic vectors. Therefore, by simple geometric reasoning, the final field from such a superposition will necessarily have the total electric and magnetic vectors aligned with the height of the tetrahedron of the particular Cartesian frame as shown in the schematic of Fig. 1.

Theoretical evidence for the existence of such configurations in hybrid modes has first been given in Shimoda and Uehara [17], [18] and they were later justified as special solutions of Maxwell equations [19]. In the case presented, the overall configuration would be easier to set up in an actual experiment by simply crossing three specially devised discharge mechanisms along each of the three axes in a Cartesian system in the lab frame.

A separate, interesting question concerns the possibility of an overall rotation of the reference frame upon which the electric and magnetic vectors are located. There are two cases of special interest, one being that of the corresponding angular momentum vector being parallel to the diagonal where the electric and magnetic vectors reside and the other concerns the case of the angular momentum vector being normal to any plane containing the said diagonal. Combination of the two is of course a more general possibility. Treatment of such a problem in the short space of this report is not possible and may be given in a subsequent article.

Last but not least, is should be mentioned that the use of the particular Whittaker scalars was chosen only for their interesting internal symmetry of the resulting vector algebra. One could introduce for example a simpler harmonic dependence by choosing arbitrary functions $F(z)\cos(\omega t), G(x,y), \psi(x,y)$ although not necessarily with exactly the same vector algebra. Nevertheless the particular property of parallelism of electric and magnetic vectors from the 9 occurring independent fields under the action of the $\hat{\pi}$ operator could still hold true. Hence, the author believes that this particular technique offers significant advantages in the search for unusual electromagnetic configurations that are difficult to treat analytically by other existing methods.

## 5. References

[1] E. T. Whittaker, "*On an Expression of the Electromagnetic Field Due to Electrons by means of Two Scalar Potential Functions*", Proc. London Math. Soc. **1** (1904) 367

[2] H. Bateman, "*The Solution of Partial Differential Equations by means of Definite Integrals* ", Proc. London Math. Soc. **1** (1904) 451

[3] H. S. Ruse, *"The geometry of the electromagnetic six-vector, the electromagnetic energy and the Hertzian tensor"*, C. R. Congr. internat. Math., **2** (1936) 232.

[4] H. S. Ruse, *"On Whittaker's Electromagnetic Scalar Potentials"*, Quart. J. Math. Soc. **8** (1937) 148

[5] H. Kawaguchi, S. Murata, *"Hertzian Tensor Potential Which Results in Lienard-Wiechert Potential "*, J. Phys. Soc. Jap. **58**(3) (1989) 848

[6] H. Kawaguchi, T. Honma, *"On the Super-Potentials for Lienard-Wiechert Potentials in Far Fields"*, J. Phys. **A**: Math. Gen. **25** (1992) 4437

[7] H. Kawaguchi, T. Honma, *"Superpotentials of Lienard-Wiechert Potentials in far fields: The Relativistic Case"*, J. Phys. **A**: Math. Gen. **26** (1993) 4431

[8] H. Kawaguchi, T. Honma, *"On a Double Fiber Bundle Structure of the Lienard-Wiechert Superpotentials"*, J. Tech. Phys. **35**(1-2) (1994) 61

[9] H. Kawaguchi, T. Honma, *"On the Electrodynamics of the Lienard-Wiechert Superpotentials"*, J. Phys. **A**: Math. Gen. **28** (1995) 469

[10] H. Marmanis, PhD Thesis, *"Analogy between the Electromagnetic and Hydrodynamic Equations: Applications to Turbulence"*, Brown University, 1999.

[11] A. A. Martins, M. J. Pinheiro, *"Fluidic electrodynamics: Approach to electromagnetic propulsion"*, Phys. Fluids **21** (2001) 097103

[12] H. Bateman, *"Partial Differential Equations of Mathematical Physics"*, Cambridge Univ. Press, 1959

[13] D. P. Stern, *"Euler Potentials"*, Am. J. Phys. **38**(4), (1970) 494

[14] G. S. Asanov, *"Clebsch Representations and Energy-Momentum of Classical Electromagnetic and Gravitational Fields"*, *Found. Phys.* **10**(11/12), (1980)

[15] J. Marsden, A. Weinstein, *"Coadjoint Orbits, Vortices and Clebsch Variables for Incompressible Fluids"*, Physica **7**D, (1983) 305

[16] A. F. Ranada, "*Interplay of Topology and Quantization: Topological Energy Quantization in a Cavity*", Phys. Let. **A** 310 (2003) 434

[17] K. Uehara *et al.*, "*Non-Transverse Electromagnetic Fields with Parallel Electric and Magnetic Fields*", J. Phys. Soc. Jap. **58**(10), (1989) 3570

[18] K. Shimoda *et al.*, "*Electromagnetic Plane Waves with Parallel Electric and Magnetic Fields E‖B in Free Space*", Am. J. Ph*ys*. **58**(4), (1990) 394

[19] J. E. Gray, "*Electromagnetic Waves with E Parallel to B*", J. Phys. **A:** Math. Gen.*, **25**, (1992) 5373

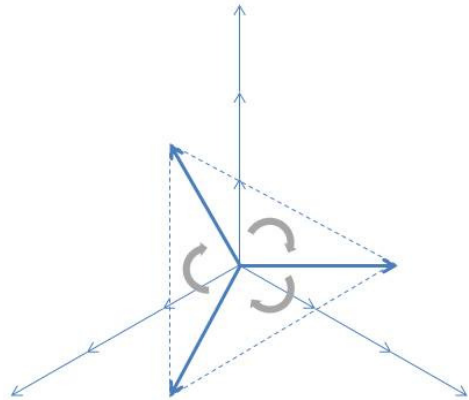

**Fig. 1.** A schematic showing the symmetric tetrahedron formed by the permutation of the coordinates of either the electric or the magnetic field components. Their common summand must then lie on the height of the tetrahedron drawn from the origin of the coordinate system.